\makeatletter
\@ifundefined{@parse@version@dash}{%
\def\@parse@version#1{\@parse@version@0#1}
\def\@parse@version@#1/#2/#3#4#5\@nil{%
\@parse@version@dash#1-#2-#3#4\@nil}
\def\@parse@version@dash#1-#2-#3#4#5\@nil{%
  \if\relax#2\relax\else#1\fi#2#3#4 }
}{}
\makeatother

\documentclass[%
reprint,
superscriptaddress,
nofootinbib,
 amsmath,amssymb,
 aps,
prl,
floatfix,
]{revtex4-2}

\usepackage{dcolumn}
\usepackage{bm}

\usepackage{aas_macros}
\usepackage{xspace}
\usepackage{color}
\usepackage{hyperref}

\begin{document}

\preprint{APS/123-QED}

\title{Comment on ``No Black Holes from Light'', PRL{\bf 130}, 041401 (2024)}

\author{Abraham Loeb}
\email{aloeb@cfa.harvard.edu}

\affiliation{Department of Astronomy, Harvard University, 60 Garden Street, Cambridge, MA 02138, USA}

\date{\today}

\begin{abstract}
We show that black holes can be made of light by adding gravity to the
discussion of Alvarez-Dominguez et al. [PRL{\bf 130}, 041401 (2024)].
\end{abstract}

\maketitle

Recently, Alvarez-Dominguez et al. published a paper titled: “No Black
Holes from Light”~\citep{2024PhRvL.133d1401A}. The authors did their
calculations in the Minkowski metric, ignoring gravity. Here we
demonstrate that gravity changes their conclusions.

Early in the history of the Universe, the cosmic energy budget was
dominated by radiation composed of photons and relativistic particles,
including electron-positron pairs, neutrinos of all flavors, $\mu$ and
$\tau$ leptons and anti-leptons, W$+$, W$-$ and Z bosons, and a plasma
of quarks and gluons of all types in thermal equilibrium. The Universe
expanded slowly enough for these particles to reach thermodynamic
equilibrium in the first 10-microseconds after the Big-Bang. In this
thermal state, the process considered by the authors - pair creation
by vacuum polarization, also known as the Schwinger
effect~\citep{1951PhRv...82..664S}, was balanced by the reverse
process of pair annihilation, not only for electron-positron pairs but
for all particle-antiparticle pairs. The existence of relativistic
particles, in addition to photons in thermal equilibrium, only changes
the effective number of degrees of freedom in the energy density of
the radiation field at a given temperature~\citep{2024pca..book.....K}
that shapes the evolution of spacetime under gravity through the
associated energy-momentum tensor.

Einstein’s equations of General Relativity imply that an expanding
spherical region in the early Universe, filled with radiation in
thermal equilibrium of an energy density above the critical value,
$3H^2/8\pi G$, where $H(t)$ is the Hubble parameter at time $t$, would
collapse to a black
hole~\citep{1974MNRAS.168..399C,2024arXiv240605736C,2024arXiv240605736C}.
If the overdensity is of order unity on the Hubble scale, $c/H$, the
collapse occurs on a timescale of order a few $t$.  The spherical
region, whose interior is described by the
Friedmann-Lemaire-Robertson-Walker spacetime, could have a sharp outer
boundary and be surrounded by vacuum. In that case, the spacetime
outside of this boundary would be described by the Schwarzschild
solution for a black hole with a mass $M$ equal to the total
mass-energy enclosed within the finite sphere. Applying Birkhoff’s and
Israel's theorems~\citep{1967PhRv..164.1776I} to the
spherically-symmetric spacetime interior to the boundary, implies that
it follows the evolution of a closed Universe that eventually
collapses to a singularity within a finite time in a Big Crunch. This
final state would be a black hole, described by the full Schwarzschild
spacetime.

It is well known that in order to create a black hole, one needs to
concentrate a mass $M$ within a Schwarzschild radius of $\sim
2GM/c^2$, as formulated by the so-called “Hoop
Conjecture”\citep{1983CMaPh..90..575S}. Any choice of black hole mass,
$M$, or the equivalent radiation energy $E=Mc^2$, translates to the
radius of the progenitor region that will collapse to make the black
hole.

The radius of the cosmic horizon at a time of $\sim 10\mu$s after the
Big-Bang was $\sim 3$~km. The radiation temperature at that time was
150~MeV$=294m_ec^2$, with an equivalent mass density of $\sim 2\times
10^{16}~{\rm g~cm^{-3}}$. It is possible to create a Schwarzschild
black hole with the mass of the Sun by concentrating a thermal
radiation bath at this temperature of total energy
$E=1.8x10^{47}$~Joules within a spherical region smaller than 3~km.

Finally, we note that black hole creation out of light does not demand
a radiation field of extreme intensity. If we were to fill the
present-day universe with a thermal blackbody radiation at room
temperature, $\sim 300$K, no electron-positron pairs would be produced
by the Schwinger mechanism of the associated infrared
photons. However, the radiation energy density would exceed the
critical value for collapse to a black hole singularity by a factor of
$\sim 3\times 10^4$. Currently, we do not live inside the event horizon a
black hole because the actual temperature of the cosmic microwave
background, 2.73~K, is 110 times smaller, making the
radiation field sub-critical by a factor of $\sim 5\times 10^3$.

In conclusion, black holes can be made out of light. In fact, it is
possible that dark matter is primordial black holes in the
asteroid-mass range of $\sim
10^{17}$-$10^{22}$~g~\citep{2021arXiv211002821C,2024NuPhB100316494G}. If
that is the case, most of the matter in today’s Universe was created
by the collapse of thermal radiation into black holes in the first 10
picoseconds after the Big Bang. But even today, giant black holes
could be created by concentrating low-energy photons in a large enough
volume of space.

\textbf{Acknowledgments} -- This work was supported in part by the Black Hole
Initiative which is funded by JTF and GBMF.

\bibliography{t}
\end{document}